\documentclass{article}

\usepackage{epsfig}
\usepackage{graphicx}
\usepackage{amsmath}
\usepackage{amssymb}
\usepackage{graphics}
\usepackage[left=3cm,right=3cm,top=3cm,bottom=2cm]{geometry}

\title{The Extended Nuclear Matter Model with Smooth Transition Surface}

\author{Jorge A. Rueda H.$^{1,2}$, B. Patricelli$^{1,2}$, M. Rotondo$^{1,2}$, R. Ruffini$^{1,2,3}$ and S-S. Xue$^{2}$ \\\\
$^1$ University of Rome ``La Sapienza''--Piazzale Aldo Moro 5
--00185, Rome--Italy\\
$^2$ ICRAnet and ICRA, Piazzale della Repubblica 10, 65122, Pescara--Italy\\
$^3$ ICRAnet, University of Nice ``Sophia Antipolis'', Grand Ch\^{a}teau, BP 2135, 06103 Nice--France}

\begin{document}

\maketitle

\abstract{The existence of electric fields close to their critical value $E_c=\frac{m_e^2
c^3}{e \hbar}$ has been proved for massive cores of $10^7$ up to $10^{57}$
nucleons using a proton distribution of constant density and a sharp step
function at its boundary \cite{migdal,michael,prl}.  We explore the modifications
of this effect by considering a smoother density profile with a proton
distribution fulfilling a Woods--Saxon dependence.  The occurrence of a critical
field has been confirmed. We discuss how the location of the maximum of the
electric field as well as its magnitude is modified by the smoother
distribution.}

%%%%%%%%%%%%%%%%%%%%%%%%%%%%%%%%%%%%%%%%%%%%%%%%%%%%%%%%%%%%%%%%%%%%%%%%%%%%%%%%%%
\section{Introduction}
%%%%%%%%%%%%%%%%%%%%%%%%%%%%%%%%%%%%%%%%%%%%%%%%%%%%%%%%%%%%%%%%%%%%%%%%%%%%%%%%%%

One of the most active field of research has been to analyze a general approach to Neutron Stars based on the Thomas-Fermi ultrarelativistic equations amply adopted in the study of superheavy nuclei. The aim is to have a unified approach both to superheavy nuclei, up to atomic numbers of the order of $10^5$--$10^6$, and to what we have called ``Massive Nuclear Cores''.  These cores are characterized by atomic number of the order of $10^{57}$, composed by neutrons, protons and electrons in $\beta$--equilibrium, and expected to be kept at nuclear density by self gravity.

The analysis of superheavy nuclei has historically represented a major field of research, guided by Prof. V. Popov and Prof. W. Greiner and their schools. This same problem was studied in the context of the relativistic Thomas-Fermi equation also by R. Ruffini and L.Stella, already in the '80s. A more recent approach \cite{michael,prl} has shown the possibility to extrapolate this treatment of superheavy nuclei to the case of Massive Nuclear Cores.

The very unexpected result has been that also around these massive cores there is the distinct possibility of
having an electric field close to the critical value $E_c = \frac{m_e^2 c^3}{e \hbar}$,
although localized in a very narrow shell of the order of the electron Compton wavelength.

In all the mentioned works has been assumed a sharp profile for the proton distribution as given by a step function centered on the surface of the core; so modeling a sharp transition surface between for example, the core and the crust of neutron stars.  In this work we model the transition surface in a smoother way by relaxing the sharp profile of the proton distribution and analyze the changes that it produce on the general properties of the system.

%%%%%%%%%%%%%%%%%%%%%%%%%%%%%%%%%%%%%%%%%%%%%%%%%%%%%%%%%%%%%%%%%%%%%%%%%%%%%%%%%%
\section{The Relativistic Thomas--Fermi Equation}
%%%%%%%%%%%%%%%%%%%%%%%%%%%%%%%%%%%%%%%%%%%%%%%%%%%%%%%%%%%%%%%%%%%%%%%%%%%%%%%%%%

Let us to introduce the proton distribution function $f_p(x)$ by mean of
$n_p(x) = n^c_p f_p(x)$, where $n^c_p$ is the central number density of
protons. We use the dimensionless unit $x=(r-R_c)/a$, with $a^{-1}=\sqrt{4 \pi
\alpha \lambda_e n^c_p}$, $\lambda_e$ is the electron Compton wavelength, $R_c$
the point where initial conditions are given ($x=0$) and $\alpha$ is the fine
structure constant.

Using the Poisson's equation and the equilibrium condition for the gas of
electrons

\begin{equation}\label{eqcond1}
E^e_F = m_e c^2 \sqrt{1+x^2_e}-m_e c^2-e V = 0\, ,
\end{equation}

where $e$ is the fundamental charge, $x_e$ the normalized electron Fermi
momentum and $V$ the electrostatic potential, we obtain the relativistic
Thomas--Fermi equation

\begin{equation}\label{eq:tf}
\xi_e''(x)+\left( \frac{2}{x+R_c/a} \right)
\xi_e'(x)-\frac{[\xi^2_e(x)-1]^{3/2}}{\mu}+f_p(x)=0\, ,
\end{equation}

where $\mu=3 \pi^2 \lambda^3_e n^c_p$ and we have introduced the normalized
electron chemical potential in absence of any field $\xi_e = \sqrt{1+x^2_e}$.
For a given distribution function $f_p(x)$ and a central number density of
protons $n^c_p$, the above equation can be integrated numerically with the
boundary conditions
\begin{equation}
\xi_e(0) =\sqrt{1+\left[\mu\,\delta f_p(0) \right]^{2/3}} \, ,\qquad \xi_e'(0)
< 0\, ,
\end{equation}
where $\delta \equiv n_e(0)/n_p(0)$.

After integrating the TF equation, we can to calculate the neutron number
density using the equilibrium condition of the direct and inverse $\beta$ decay

\begin{equation*}
n \rightarrow e^{-}+p+\bar{\nu}\, ,\qquad 
e^{-}+p \rightarrow n+\nu\, ,
\end{equation*}

which results in

\begin{equation}
m_n c^2 \xi_n-m_n c^2=m_p c^2 \xi_p-m_p c^2+e V
\end{equation}

The electrostatic potential $V$ is calculated using the equilibrium condition (\ref{eqcond1}).

%%%%%%%%%%%%%%%%%%%%%%%%%%%%%%%%%%%%%%%%%%%%%%%%%%%%%%%%%%%%%%%%%%%%%%%%%%%%%%%%%%
\section{The Woods-Saxon--like Proton Distribution Function}
%%%%%%%%%%%%%%%%%%%%%%%%%%%%%%%%%%%%%%%%%%%%%%%%%%%%%%%%%%%%%%%%%%%%%%%%%%%%%%%%%%

\begin{figure}
\centering
\includegraphics[scale=0.4]{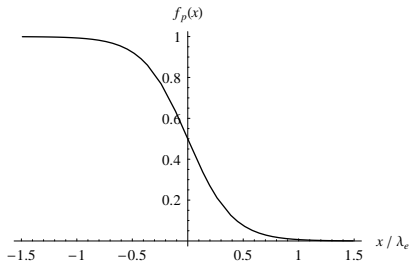}
\caption{Proton distribution function for $\gamma=1.5$, $\beta \approx
0.0585749$.}\label{fig:fp}
\end{figure}

We propose a monotonically decreasing proton distribution function fulfilling
a Woods--Saxon dependence
\begin{equation}\label{eq:fp}
f_p(x) = \frac{\gamma}{\gamma+e^{\beta x}}\, ,
\end{equation}
where $\gamma > 0$ and $\beta > 0$.  In fig. \ref{fig:fp} we show the proton
distribution function for a particular set of parameters.

%%%%%%%%%%%%%%%%%%%%%%%%%%%%%%%%%%%%%%%%%%%%%%%%%%%%%%%%%%%%%%%%%%%%%%%%%%%%%%%%%%
\section{Results of the Numerical Integration}
%%%%%%%%%%%%%%%%%%%%%%%%%%%%%%%%%%%%%%%%%%%%%%%%%%%%%%%%%%%%%%%%%%%%%%%%%%%%%%%%%%

We have integrated numerically the eq.(\ref{eq:tf}) for several sets of
parameters and initial conditions.  To show the general behaviour of the system we show here the following two samples of initial conditions

\begin{table}
\centering
\begin{tabular}{|c|c|c|}
%\begin{equation*}
%\begin{array}{|c|c|c|}
\hline
$\delta$ & $\xi'_e(0)$ & $n^c_p (cm^{-3})$\\
\hline
0.9662053 & -0.8680512263367902 & $1.38\times 10^{36}$\\
0.97829293547 & -0.899201 & $2.76\times 10^{36}$\\ \hline
%\end{array}
\end{tabular}
\caption{Sets of initial conditions}\label{table:conditions}
%\end{equation*}
\end{table}

for the which we obtain respectively the physical quantities

\begin{table}
\centering
\begin{tabular}{|c|c|c|c|}
%\begin{equation*}
%\begin{array}{|c|c|c|c|}
\hline
$N_e=N_p$ & $A$ & $E_{peak}/E_c$ & $R_c (km)$\\
\hline
$10^{54}$ & $1.61\times 10^{56}$ & 95 & 5.56\\
$2\times 10^{54}$ & $2.35\times 10^{56}$ & 125 & 5.56\\ \hline
%\end{array}
\end{tabular}
\caption{Physical quantities for the sets of parameters in table \ref{table:conditions}}
%\end{equation*}
\end{table}

%-----------------------------------------------------------------------------
\subsection{Number Densities}
%-----------------------------------------------------------------------------

\begin{figure}
\centering
\includegraphics[scale=0.45]{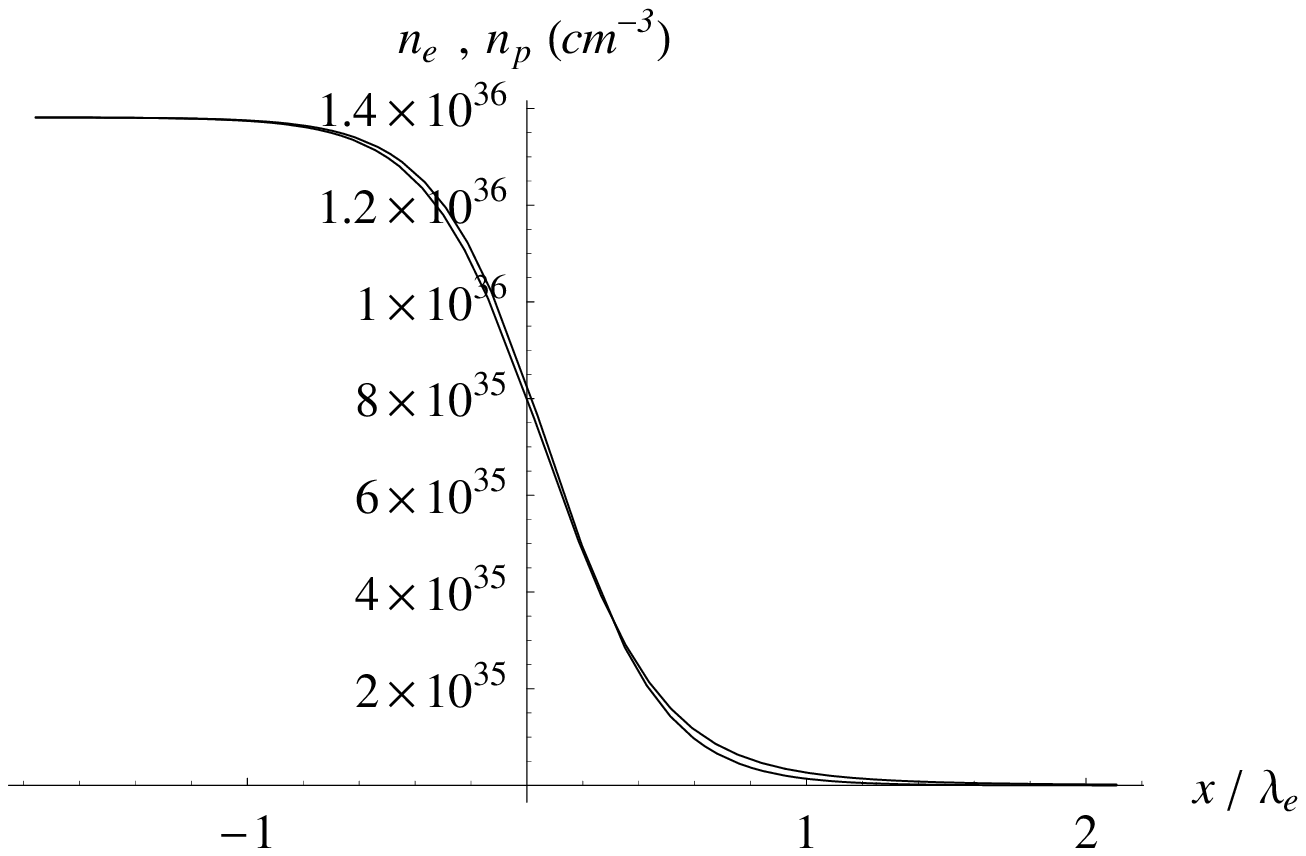} \includegraphics[scale=0.45]{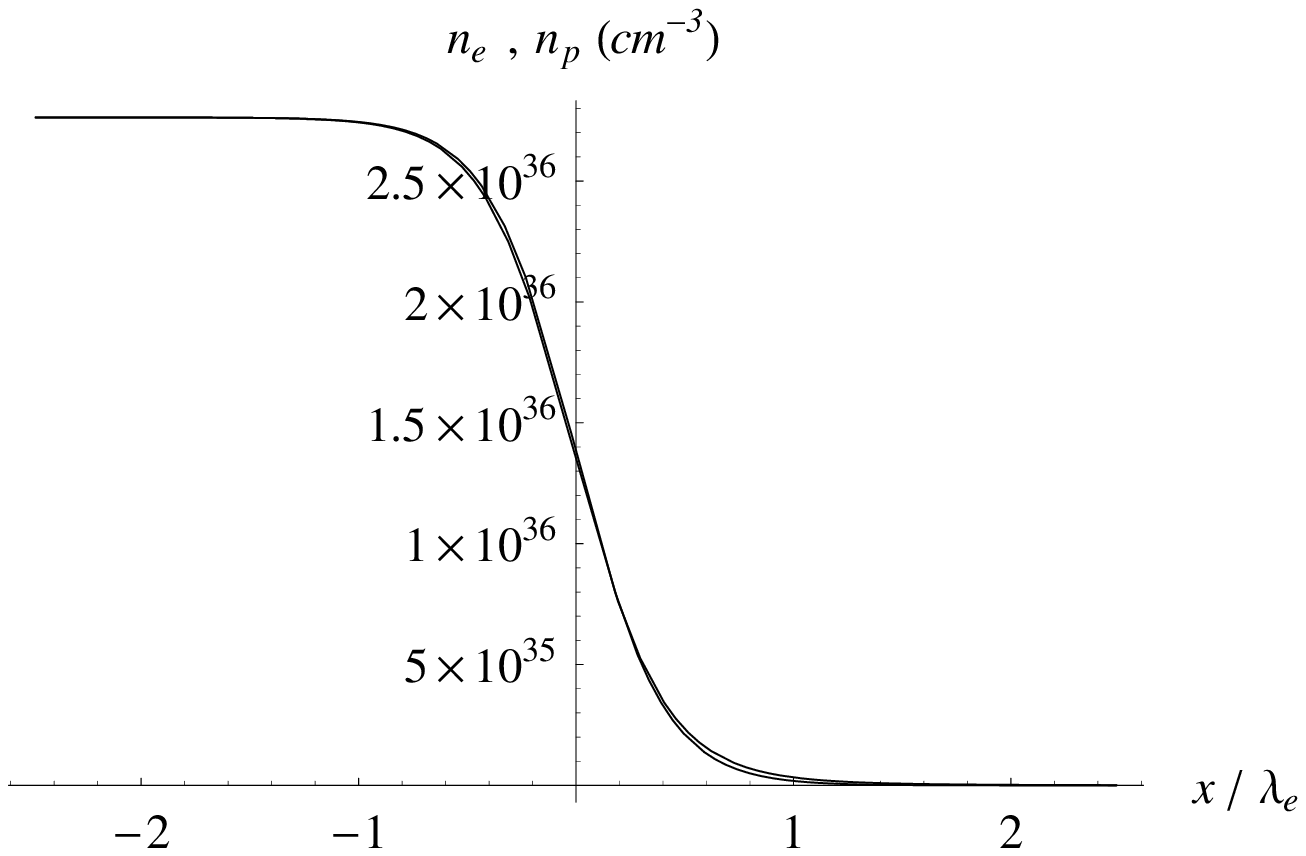}
\caption{Electron and Proton Number Density for the sets of parameters in table \ref{table:conditions} }\label{fig:densities}
\end{figure}

In fig.\ref{fig:densities} we show the electron and proton number density together.  We can see here that there is no local charge neutrality because the small difference between the two profiles.  Nevertheless, this small difference creates a charge separation (see fig.\ref{fig:chseparation}) enough to produce huge electric fields (see fig.\ref{fig:fields}).

We can also see how the system reaches indeed global charge neutrality in a very small scale as noted by Migdal \emph{et al.} \cite{migdal} in their classical paper.  This scale have been calculated to be of the order of $\lambda_{\pi}/\sqrt{\alpha}$.

%-----------------------------------------------------------------------------
\subsection{The Charge Separation}
%-----------------------------------------------------------------------------

\begin{figure}
\centering
\includegraphics[scale=0.45]{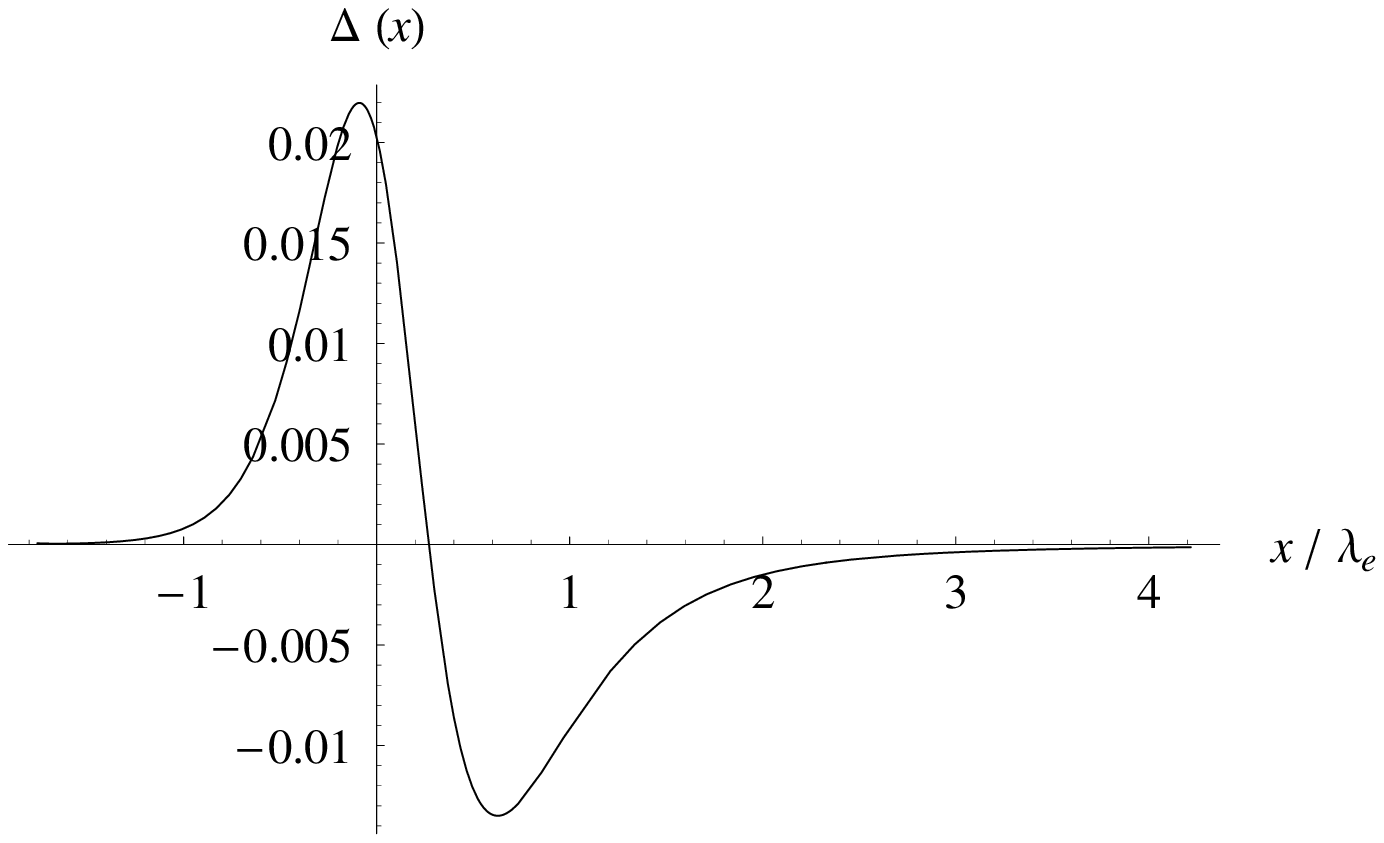} \includegraphics[scale=0.45]{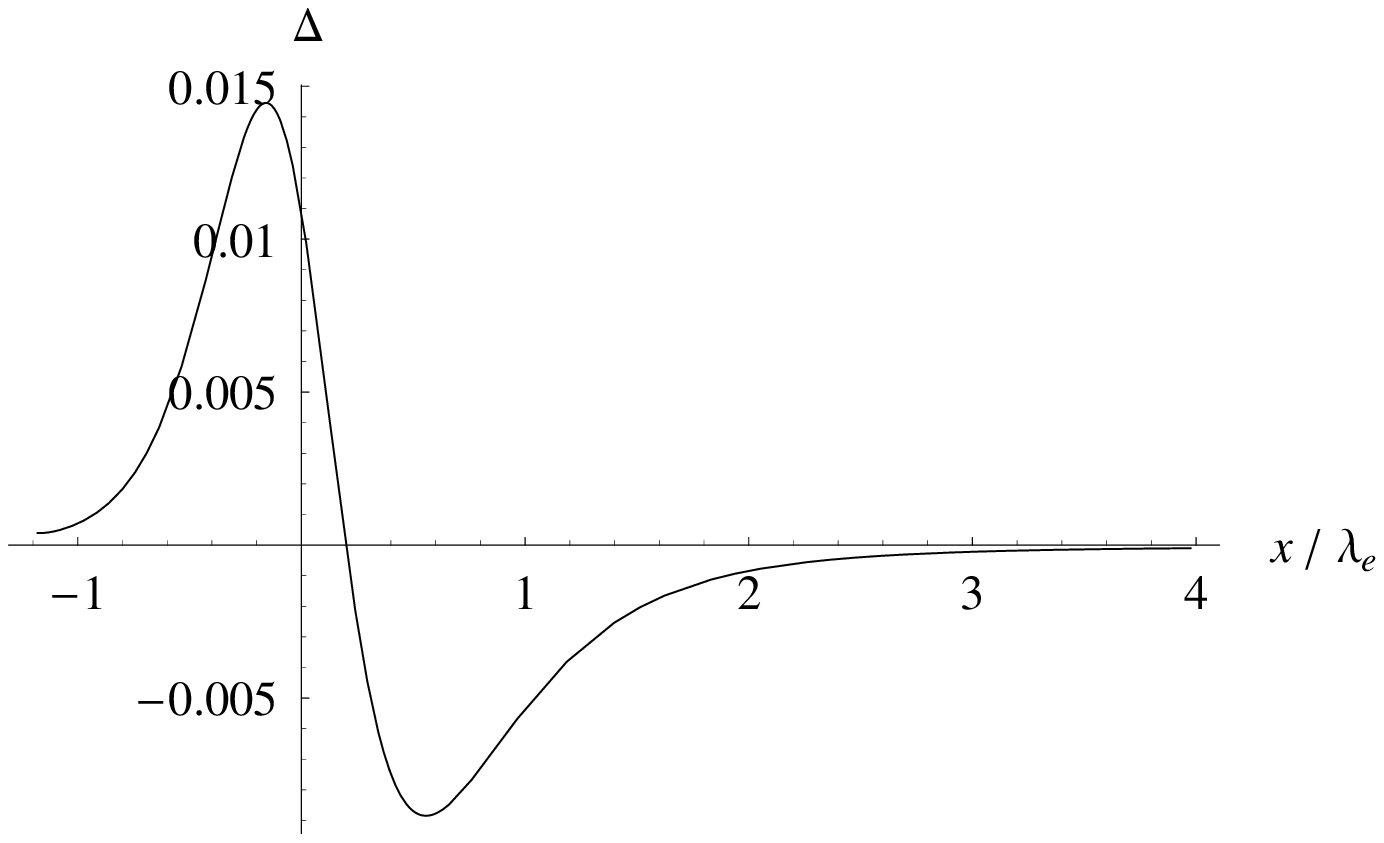}
\caption{Charge separation for the sets of parameters in table \ref{table:conditions} }\label{fig:chseparation}
\end{figure}

In order to see more clearly the difference between the electron and the proton profiles we have plotted in fig.\ref{fig:chseparation} the charge separation function given by

\begin{equation}
\Delta(x)=\frac{n_p(x)-n_e(x)}{n^c_p}\, .
\end{equation}

We see two well defined zones with opposite charge.  In the first zone we have $n_p>n_e$ so we have a positive charged shell while in the second one we have $n_p<n_e$ and a negative charged shell develops.  At the point $n_e=n_p$ we have a maximum of the electric field, which is screened by the negative charged shell until reach global charge neutrality (see fig.\ref{fig:fields}).

%-----------------------------------------------------------------------------
\subsection{The Electric Field}
%-----------------------------------------------------------------------------

\begin{figure}
\centering
\includegraphics[scale=0.45]{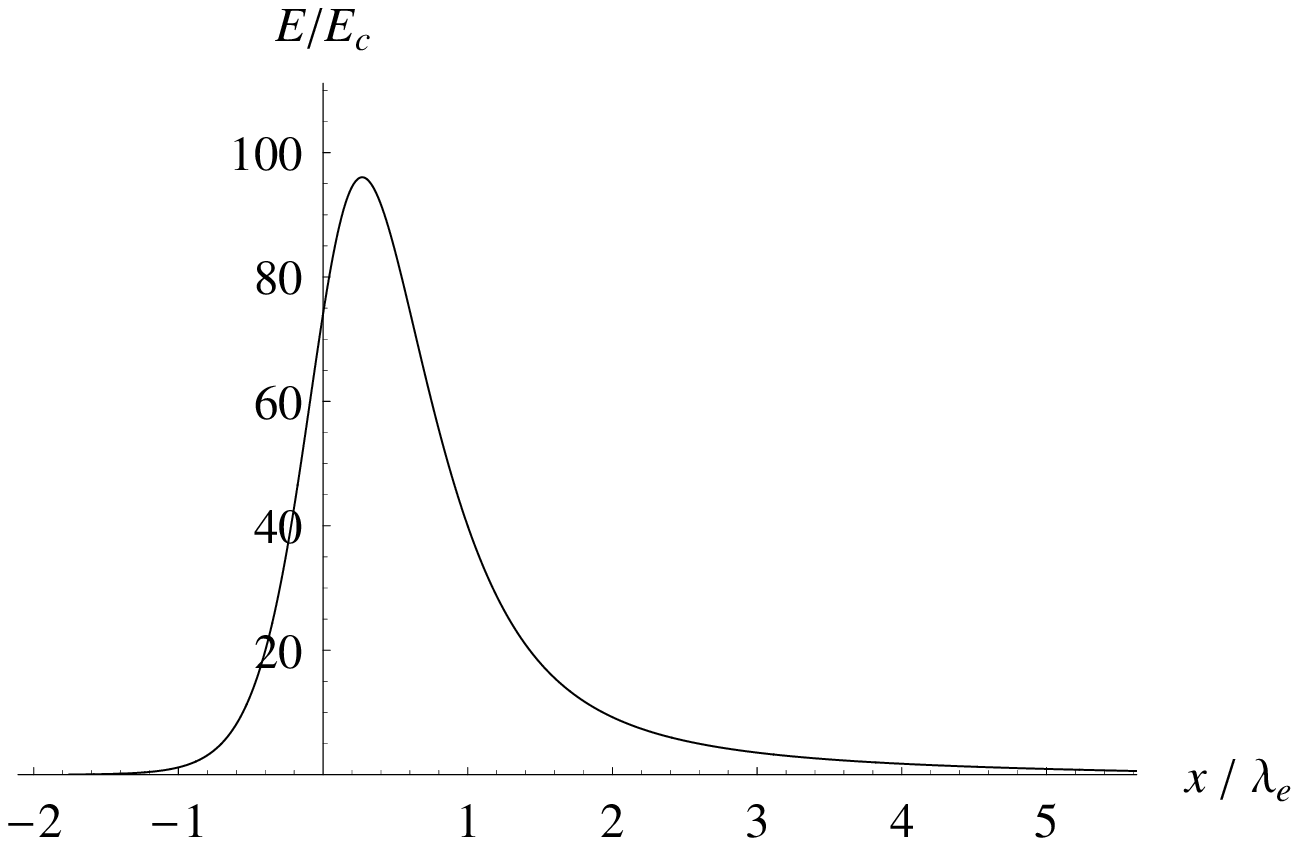} \includegraphics[scale=0.45]{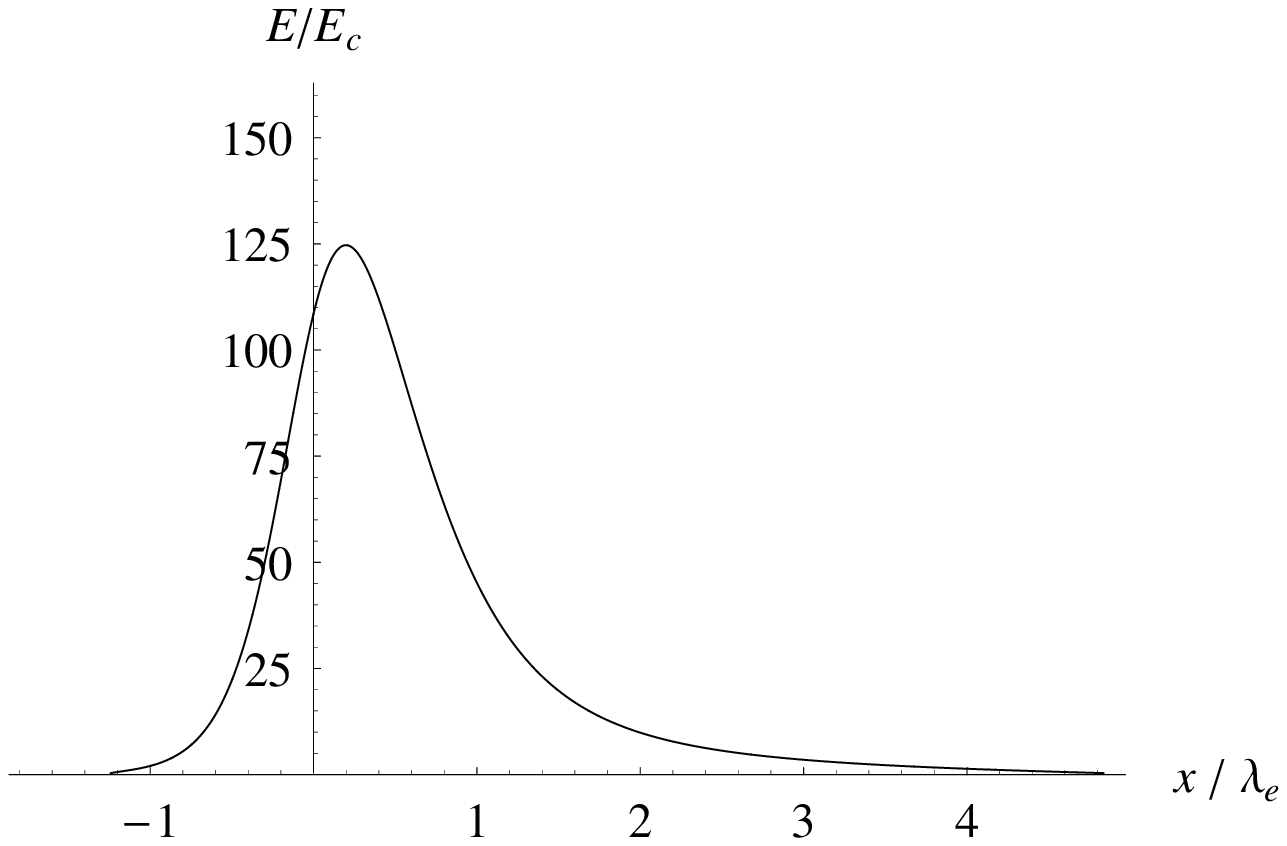}
\caption{Electric Field for the sets of parameters in table \ref{table:conditions} }\label{fig:fields}
\end{figure}

We have plotted in fig.\ref{fig:fields} the electric field in units of the critical field $E_c = \frac{m_e^2 c^3}{e \hbar}$, namely

\begin{equation}
\frac{E}{E_c}=-\frac{\lambda_e}{a} \xi'_e(x)\, .
\end{equation}

We see that the electric field is overcritical but smaller in respect to the case
of a sharp step proton distribution as used in \cite{migdal,michael,prl}.  Nevertheless, it is yet well above the critical field $E_c$.

The maximum of the electric field occur at the point where the transition from the positive charged shell to the negative charged one takes place, i.e., where $n_e=n_p$ (see fig.\ref{fig:chseparation}).  Of course because we have assumed a smoother proton profile we find also a smoother electric field about the maximum.  We recall that in the case of sharp proton profile the first derivative of the electric field has a discontinuity on the point of charge inversion.

%%%%%%%%%%%%%%%%%%%%%%%%%%%%%%%%%%%%%%%%%%%%%%%%%%%%%%%%%%%%%%%%%%%%%%%%%%%
\section{Conclusions}
%%%%%%%%%%%%%%%%%%%%%%%%%%%%%%%%%%%%%%%%%%%%%%%%%%%%%%%%%%%%%%%%%%%%%%%%%%%

We confirm the existence of overcritical electric fields in the smooth transition surface of a massive nuclear core.  The intensity of the electric field depends on the proton density mainly by two factors: the first one is the value of $n_p$ about the surface and the second one is how it changes about the surface (sharpness).  In this line we note that the first factor depends strongly on a precise value of the so called `melting density' and the correct value of the charge to mass ratio ($Z/A$) as given by $\beta-$equilibrium, while for the second factor could be very important the surface tension as given for instance by the strong interaction.

%%%%%%%%%%%%%%%%%%%%%%%%%%%%%%%%%%%%%%  References %%%%%%%%%%%%%%%%%%%%%%%%%%%%%%%%


\begin{thebibliography}{99}

\bibitem{migdal} A. B. Migdal, D. N. Voskresenskii and V. S. Popov, {\it JETP
letters}, {\bf 24} 186 (1976)

\bibitem{michael} R. Ruffini, M. Rotondo and S. S. Xue, {\it Int. Journal of Mod. Phys
D}, {\bf 16} 1 (2007)

\bibitem{prl} V. S. Popov, M. Rotondo, R. Ruffini, and S. S. Xue, {\it In preparation.} (2009)

\end{thebibliography}
\end{document}